\documentclass[nofootinbib,a4paper,aps,twocolumn,prd,10pt,superscriptaddress]{revtex4}

\usepackage{graphicx}
\usepackage{amsmath}
\usepackage{amsfonts}
\usepackage{amsthm}
\usepackage{mathrsfs}
\usepackage{mathtools}
\usepackage{amssymb}
\usepackage{epsfig}
\usepackage{amscd}
\usepackage{dcolumn}
\usepackage{bm}
\usepackage{natbib}
\usepackage{url}
\usepackage{xspace}
\usepackage{kantlipsum}

\usepackage[normalem]{ulem}
\usepackage{appendix}
\usepackage{algorithm}
\usepackage{algorithmic}
\usepackage{hyperref}
\usepackage{nameref}

\usepackage{siunitx}
\usepackage{paralist}
\usepackage{placeins}

\usepackage{xcolor}

\usepackage{comment}

\usepackage{hyperref}
\hypersetup{%
    ,urlcolor=blue
    ,citecolor=blue
    ,linkcolor=blue
    }

\def\be{\begin{equation}}
\def\ee{\end{equation}}
\def\bea{\begin{eqnarray}}
\def\eea{\end{eqnarray}}

\begin{document}

\title{Constraining quadrupole deformations with 
 relativistic effects}

\author{Daniya~\surname{Utepova}}
\email[]{utepova\_daniya@mail.ru}
\affiliation{Abai Kazakh National Pedagogical University, Dostyk Ave., 13, Almaty, 050010, Kazakhstan}
\affiliation{Al-Farabi Kazakh National University, Al-Farabi Ave., 71, Almaty, 050040, Kazakhstan}

\author{Kuantay~\surname{Boshkayev}}
\email[]{kuantay@mail.ru}
\affiliation{Al-Farabi Kazakh National University, Al-Farabi Ave., 71, Almaty, 050040, Kazakhstan}
\affiliation{Institute of Nuclear Physics, Ibragimova, 1, Almaty, 050032, Kazakhstan}

\author{Serzhan~Momynov}
\email[]{s.momynov@gmail.com}
\affiliation{Al-Farabi Kazakh National University, Al-Farabi Ave., 71, Almaty, 050040, Kazakhstan}
\affiliation{Satbayev University, Satbayev St., 22, Almaty, 050013, Kazakhstan}

\author{Anuar~\surname{Idrissov}}
\email[]{anuar.idrissov@gmail.com}
\affiliation{Department of Physics, Nazarbayev University, Kabanbay Batyr 53, 010000 Astana, Kazakhstan}
\affiliation{Fesenkov Astrophysical Institute, Observatory 23, 050020 Almaty, Kazakhstan}

\author{Gulzada~\surname{Baimbetova}}
\email[]{bgulzada_74@mail.ru}
\affiliation{Abai Kazakh National Pedagogical University, Dostyk Ave., 13, Almaty, 050010, Kazakhstan}

\author{Hernando~\surname{Quevedo}}
\email[]{quevedo@nucleares.unam.mx}
\affiliation{Instituto de Ciencias Nucleares, Universidad Nacional Aut\'onoma de M\'exico, Mexico}
\affiliation{Dipartimento di Fisica and ICRA, Universit\`a di Roma “La Sapienza”, Roma, Italy}
\affiliation{Al-Farabi Kazakh National University, Al-Farabi Ave., 71, Almaty, 050040, Kazakhstan}

\author{Ainur~\surname{Urazalina}}
\email[]{y.a.a.707@mail.ru}
\affiliation{Al-Farabi Kazakh National University, Al-Farabi Ave., 71, Almaty, 050040, Kazakhstan}
\affiliation{Institute of Nuclear Physics, Ibragimova, 1, Almaty, 050032, Kazakhstan}

\date{\today}

\begin{abstract}

We investigate two general relativistic effects -- namely, the Shirokov and Shapiro effects -- within the framework of the Zipoy--Voorhees spacetime ($q$--metric), which generalizes the Schwarzschild solution by incorporating a quadrupole moment. By analyzing the geodesic deviation equations, we explore the oscillatory motion of test particles and demonstrate how the source's quadrupole parameter influences the Shirokov effect. Furthermore, we derive an expression for the Shapiro time delay in this deformed spacetime and examine the quadrupole moment's impact on the gravitational time delay experienced by radio waves propagating near a massive object. 
The first--order approximation reveals a pronounced effect of the quadrupole parameter on the time delay, in contrast to similar recent analyses. 
These findings deepen our understanding of how deviations from spherical symmetry influence gravitational phenomena, with potential implications for the study of compact astrophysical objects such as neutron stars and naked singularities or ''black hole mimickers'' that exhibit significant multipolar structures.
\end{abstract}

\maketitle 

\section{Introduction}

Since the formulation of General Relativity (GR) in 1915 by Albert Einstein, the theory has passed numerous tests with remarkable accuracy \cite{einstein1915feldgleichungen}. The main classical experimental confirmations include the perihelion precession of Mercury, the deflection of light by gravity, gravitational redshift, and the direct detection of gravitational waves \cite{will2014confrontation,dyson1920ix,pound1960apparent,abbott2016observation}.

Among the theoretical frameworks within GR, the study of geodesic deviation is essential for understanding the relative motion of test particles in curved spacetime, making it a powerful tool for probing theoretical effects predicted by GR in various spacetime backgrounds \cite{misner1973gravitation,wald2010general}. The impact of geodesic deviation on satellite motion was examined in \cite{philipp2015geodesic}, which proposes methods for computing orbital dynamics within the framework of general relativity. Satellite missions such as GRACE-Follow-On enable the study of Earth’s global gravitational structure by monitoring variations in the relative distances between satellites \cite{tapley2019contributions}. Accounting for relativistic effects is essential, as even small deviations from classical predictions can significantly affect measurement accuracy.

One specific application of geodesic deviation in the Schwarzschild metric is the theoretical prediction of the Shirokov effect \cite{shirokov1973one}. It consists of the following: if, inside a satellite freely rotating in circular orbit (without drift), a test mass is given small velocity in the radial direction (relative to Earth) or in a direction perpendicular to the orbital plane, then, relative to the satellite, one can observe small oscillations  in two mutually perpendicular directions, but with different frequencies. This frequency difference has a general relativistic nature and depends on the radius of the circular orbit and the gravitational radius of the Earth. The analysis of this effect showed that the oscillation frequency in the plane perpendicular to the satellite's orbit coincides with the frequency of the satellite’s orbital motion, while the frequency in the radial direction corresponds to the frequency of motion of an object along the rosette-shaped trajectory in the Schwarzschild metric \cite{vladimirov1982reference}.  Thus, the Shirokov effect is essentially another manifestation of the perihelion shift effect (in this case, perigee) of the test mass's orbit near Earth \cite{philipp2015geodesic}. 

A further development of this idea was proposed in \cite{nduka1977shirokov}, where the Shirokov effect was calculated in the Lense-Thiring spacetime, discovering an additional phase shift of oscillations caused by the angular momentum of the source of the gravitational field. In the axially-symmetric rotating Kerr metric, this effect was studied in \cite{vladimirov1981small} with the help of a monadic method for defining reference systems. 

In the post-Newtonian approximation \cite{melkumova1990calculation}, the authors analyzed the oscillatory motion of a particle relative to a non-rotating satellite moving along a circular, drift-free equatorial geodesic in a static, axisymmetric external gravitational field where an additional reflection symmetry about the equatorial plane was assumed to ensure that the geodesic remains planar.

In addition to the relativistic effect on small orbital oscillations, another key phenomenon is the Shapiro time delay — the fourth test of GR \cite{shapiro1964fourth}. It refers to the extra time a light signal takes when passing near a massive object like the Sun. In Shapiro’s experiment, a signal sent from Earth to Venus and back during superior conjunction (when both planets align with the Sun) was delayed. This delay confirms that the Sun’s gravity curves spacetime, slowing the signal compared to flat space \cite{d1992introducing}. 

Shapiro's calculations were based on a first-order expansion of the Schwarzschild metric, which matches experimental measurements \cite{bertotti2003test} . To account for rotating bodies, Dymnikova \cite{dymnikova1986gravitational} used the Kerr metric. Further improvements include higher-order mass terms \cite{he2016second}, moving black holes \cite{he2016second} and for the Generalized Uncertainty Principle \cite{okcu2021observational}. Feng and Huang \cite{feng2020optical} showed the time delay can also be modeled optically.  Additionally, measuring the time delay of quasar light as it passes near Jupiter has been proposed as a method to determine the speed of gravity \cite{kopeikin2001testing,kopeikin2003post,fomalont2003measurement} with subsequent studies offering improved accuracy and important clarifications \cite{will2003propagation}. Today, the Shapiro delay aids in pulsar timing \cite{kaspi1994high,stairs1998measurement,ben2022propagation,camilo1994high}, stellar mass estimates \cite{fonseca2021refined}, probing ultralight dark matter models \cite{hook2018probing,poddar2021constraints}, test of a black hole model within the framework of loop
quantum gravity (LQG) \cite{chen2024constraints}, in gravitational lensing of black
holes enclosed by dark matter halos \cite{qiao2024time}.

In \cite{junior2023possible}, the authors propose using the Shapiro time delay as a tool to distinguish between different black hole solutions. In this context, the results of \cite{Chakrabarty:2022fbd} are particularly interesting, as the authors study the Shapiro delay in an extended Schwarzschild solution with quadrupole moment known as the $\gamma$-metric, which includes a deformation parameter. Surprisingly, they find that the Shapiro time delay is insensitive to this deformation parameter - a result that appears counterintuitive.


Building on the preceding discussion, we adopt the Zipoy–Voorhees spacetime \cite{voorhees1970static,zipoy1966topology} as the background geometry for analyzing both the Shirokov effect and the Shapiro time delay. The Zipoy-Voorhees spacetime, often referred to as the $q-$metric, $\delta-$metric or $\gamma-$metric, is a static and axially symmetric vacuum solution of Einstein’s field equations that generalize the Schwarzschild solution by incorporating a quadrupole moment, which describes the deformation of the mass distribution \cite{quevedo2011exterior}. This metric provides a useful framework for studying deviations from spherical symmetry and the impact of quadrupole deformations on gravitational effects \cite{boshkayev2016motion,herrera2000geodesics}. 

It should be emphasized that this spacetime possesses a curvature singularity at the null surface corresponding to the black hole horizon in the case of vanishing quadrupole, thereby making the spacetime structure significantly distinct from that of the Schwarzschild solution \cite{Quevedo:2010mn}. Over the years, the $q-$ metric has been extensively investigated across a variety of contexts, including the dynamics of test particles \cite{Herrera:1998rj, Chowdhury:2011aa, Capistrano:2019qdv}, models of interior solutions \cite{Stewart:1982, Herrera:2004ra}, the study of photon spheres and black hole shadows \cite{Abdikamalov:2019ztb, Arrieta-Villamizar:2020brc, Shaikh:2022ivr, Turimov:2023lbn}, spinning particle motion \cite{Toshmatov:2019bda}, the behavior of charged particles \cite{Benavides-Gallego:2018htf, Faraji:2021vid}, particle collisions \cite{Malafarina:2020kmk}, and quasiperiodic oscillations \cite{Toshmatov:2019qih, Benavides-Gallego:2020fri,2024MNRAS.531.3876B}. Further investigations have covered gravitational lensing of photons and neutrinos \cite{Boshkayev:2020igc, Chakrabarty:2021bpr}, as well as the structure and behavior of accretion disks \cite{Boshkayev:2021chc, Shaikh:2021cvl}. Additionally, several generalizations of the 
$q$-metric have been explored, including rotating configurations \cite{Quevedo:1991zz, Toktarbay:2014yru, Allahyari:2019umx, Li:2022eue}, charged solutions \cite{Gurtug:2021noy}, those involving the Newman–Unti–Tamburino (NUT) parameter \cite{Halilsoy:1992zz, Narzilloev:2020qdc}, and even wormhole geometries \cite{Gibbons:2017jzk, Narzilloev:2021ygl}.

This paper is organized as follows: Section \ref{sec:dev} is devoted to introducing the background spacetime and deriving the geodesic deviation equations. In Section \ref{sec:shir}, we focus on the Shirokov effect and analyze small oscillations of test particles influenced by the quadrupole moment. Section \ref{sec:shap} explores the Shapiro time delay within the $q-$metric, deriving the corresponding time delay expression and discussing its dependence on the quadrupole parameter. Finally, Section \ref{sec:con}  presents the conclusions, summarizing the main results and their implications for astrophysical observations of deformed compact objects. Throughout this paper, we use natural units $G = c = 1$, and the Lorentzian signature $(-, +, +, +)$.

\section{Geodesic deviation in the \textit{q}-metric}
\label{sec:dev}

The line element of the $q$-metric can be expressed as
\begin{align} \label{metric}
     ds^{2}=-f^{1+q}dt^{2}& +f^{-q}\Bigg[g^{q(2+q)}\left( \frac{dr^{2}}{f}+r^{2}d\theta^{2}\right) + \\ \nonumber
&+r^{2}\sin^{2}\theta d\phi^{2}\Bigg],
\end{align}
where
\begin{align}
f&=f(r)=1-\frac{2m}{r}, \\ \nonumber
g&=g(r,\theta)=1+\frac{m^{2}\sin^{2}\theta}{r^{2}f},
\end{align}
where  $q$ is the deformation parameter, which characterizes the departure from spherical symmetry $(q<0)$ for prolate and oblate $(q>0)$ geometries, $m$ is the mass parameter defined via the Arnowitt-Deser-Misner (ADM) mass measured by an observer as $M = m(1+q)$.  We consider the case where the reference geodesic is a circular orbit at a fixed radial coordinate \(r\), confined to the equatorial plane \(\theta = \pi/2\).
Furthermore, we will use the following notations
\begin{align} \nonumber
r=x^{1}, \quad \theta=x^{2},\quad\phi=x^{3},\quad t=x^{4}.
\end{align}
The geodesic deviation equation is written as \cite{misner1973gravitation}
\begin{equation}\label{eq:GDE}
    \frac{D^2 \xi^{\alpha}}{d s^{2}} - R^\alpha_{\mu\nu\rho} u^{\mu} u^{\nu} \xi^{\rho} = 0,
\end{equation}
where  $D^2/d s^{2}$ is the covariant derivative, $ R^\alpha_{\mu\nu\rho}$ is the Riemann tensor, $u^{\mu}=d x^{\mu}/d s$ is the 4-velocity tangential to the geodesic and the geodesic deviation vector is given by $\xi^{\alpha}$.

Following Shirokov \cite{shirokov1973one}, one can rewrite the geodesic deviation equations in terms of the connection coefficients
\begin{equation}\label{dev1}
\frac{d^{2}\xi^{i}}{ds^{2}}+2\Gamma^{i}_{jk}u^{j}\frac{d\xi^{k}}{ds}+\frac{\partial{\Gamma^{i}_{jk}}}{\partial{x}^{l}}u^{j}u^{k}\xi^{l}=0,
\end{equation}
By substituting the non-zero Christoffel symbols one can obtain for the metric \eqref{metric} a set of equations, determining $\xi^{i}$ as a function of the proper time $s$ 
\begin{equation}\label{eq2}
\frac{d^{2}\xi^{1}}{ds^{2}}+a_1\frac{d\xi^{3}}{ds}+a_2\frac{d\xi^4}{ds}+a_3\xi^1=0,
\end{equation}
\begin{equation}\label{eq3}
\begin{gathered}
\frac{d^{2}\xi^{2}}{ds^{2}}+h \xi^2=0,
\end{gathered}
\end{equation}
\begin{equation} \label{eq4}
\begin{gathered}
\frac{d^{2}\xi^{3}}{ds^{2}}+b\frac{d\xi^1}{ds}=0,
\end{gathered}
\end{equation}
\begin{equation}\label{eq5}
\begin{gathered}
\frac{d^{2}\xi^{4}}{ds^{2}}+c\frac{d\xi^1}{ds}=0,
\end{gathered}
\end{equation}
with coefficients
\begin{equation}\label{eq6}
a_{1} = 2(m(2 + q) - r)g^{q(2 + q)}u^{3},
\end{equation}

\begin{equation}\label{eq7}
a_{2} = \frac{2m}{r^2} (1 + q) f^{1 + 2q}g^{-q(2 + q)} u^{4},
\end{equation}

\begin{equation}\label{eq8}
\begin{aligned}
a_{3} &= \frac{1}{(m - r) r^{4}} g^{q(2 + q)} \\
&\quad \times \Bigg[ 2m(1 + q) f^{2q} \frac{(r-m (1+ q))}{(r-m(3 + q))^{-1}} (u^{4})^{2} \\
&\qquad + r^{2}f^{-1} \big[ 2m^{3}q(2 + q)^{2} - 2m^{2}\big(  q(2 + q) - 1 \big)r \\
&\qquad\quad - 3mr^{2} + r^{3} \big] (u^{3})^{2} \Bigg],
\end{aligned}
\end{equation}
\begin{equation}\label{eq9}    
b = \frac{2(r - m(2 + q))u^3}{r^{2}f},
\end{equation}

\begin{equation}\label{eq10}
c = \frac{2m(1 + q)u^4}{r^{2}f},
\end{equation}

\begin{equation}\label{hcoef}
h = g^{q(2 + q)}(u^3)^{2}.
\end{equation}

For a more detailed analysis of geodesic deviation in the \(q\)-metric, including the effects of quadrupole deformation on radial and angular tidal forces, see \cite{2025EPJC...85..319I}. There, the authors compare the geodesic deviation in the \(q\)-metric to that in Schwarzschild case, examining how the deformation parameter \(q\) and  the polar angle \(\theta\) influence the stretching or compression of deviation vectors, especially near the singularity surface \(r = 2m\).

\section{Shirokov effect}
\label{sec:shir} 

In Shirokov's paper \cite{shirokov1973one}, a conceptual model for analyzing the dynamical behavior of a test particle was illustrated by considering a terrestrial satellite in the form of a hollow sphere.  The center of inertia of this hollow sphere follows a geodesic in Earth's gravitational field, which is described by the Schwarzschild solution. A small test particle is placed at the center of this sphere. At the initial moment of the satellite's proper time $s=0$, a velocity $d\xi^i/ds$ is imparted to the particle,  where $\xi^i$ denotes the 4-vector representing the particle’s displacement relative to the satellite’s center of inertia. When a small radial or vertical velocity is applied, the particles undergo small oscillations in two directions with different periods
\begin{equation}\label{difperiiods}
\Delta T = T_\theta - T_r \left(\text{or} \ T_{\phi} \right)  \approx-\frac{3m}{r}T_0,
\end{equation}
where $T_0$ is the Newtonian period.
The difference in these periods depends on the orbit radius $r$ and the gravitational radius of the central mass $m$.  This effect  is absent in Newtonian gravity and is measurable with current experimental precision.  Although the magnitude is extremely small $10^{-6}$ cm, it becomes relevant for ultra-precise satellite missions and relativistic experiments near compact objects \cite{will2014confrontation}.

To generalize this approach to axially symmetric spacetimes, we now consider the metric \eqref{metric} and the geodesic equations  
\begin{equation}\label{geodeq}
\begin{gathered}
\frac{du^{i}}{ds}+\Gamma^{i}_{jk}u^{j}u^{k}=0
\end{gathered}
\end{equation}
with $u^{1}=dr/ds=0$ and $u^{2}=d\theta/ds=0$. Assuming circular motion in the equatorial plane, only the components \( u^3 \) and \( u^4 \) of the 4-vector velocity are non-zero. Substituting into the geodesic equation and using the non-vanishing Christoffel symbols in the $ \phi- $component, we obtain 
\begin{align}\label{explgeodeq}
\frac{m(1+q)}{r^{2}} f^{1+2q} (u^{4})^{2}-(r-m(2+q))(u^{3})^{2} &=0,
\end{align}
which provides a dynamical relation between $ u^3 $ and $ u^4 $.
Additionally, the normalization condition \( g_{\mu \nu} u^\mu u^\nu = -1 \) yields 
\begin{align} \label{normeq}
f^{1+q}(u^{4})^{2}-f^{-q}r^2(u^{3})^{2}&=1,
\end{align}
offering a second independent constraint on the same variables.
%

From the above system of equations, we obtain
\begin{align}\label{u3}
(u^{3})^2= \frac{m(1 + q) f^{q}}{r^{2} (r-m(3+2q))},
\end{align}
\begin{align}\label{u4}
(u^{4})^{2}=\frac{(r-m(2+q)) f^{-(q+1)}}{(r-m(3+2q))}.
\end{align}

Substituting \eqref{u3} in \eqref{hcoef},  one can get $h=\Omega^2$, where
\begin{equation}\label{bigfreq}
\begin{gathered}
\Omega^{2}=\frac{m(1+q)}{r^3} \left(1-\frac{m(3+2q)}{r}\right)^{-1}f^{q}g^{q(2+q)}
\end{gathered}
\end{equation}
Then, Eq. \eqref{eq3} becomes
\begin{equation}\label{eq18}
\begin{gathered}
\frac{d^{2}\xi^{2}}{ds^{2}}+\Omega^{2}\xi^{2}=0,
\end{gathered}
\end{equation}
whose solution is
\begin{equation}\label{eq19}
\begin{gathered}
\xi^{2}=\xi_{0}^{2} e ^{i\Omega s}.
\end{gathered}
\end{equation}

Furthermore, the solutions of Eqs. \eqref{eq2}, \eqref{eq4}, and \eqref{eq5}  can be written in the following form
\begin{equation}\label{eq20}
\begin{gathered}
\xi^{1}=\xi_{0}^{1} e ^{i\omega s},~\xi^{3}=\xi_{0}^{3} e ^{i\omega s},~\xi^{4}=\xi_{0}^{4} e ^{i\omega s}.
\end{gathered}
\end{equation}
Substituting
\eqref{eq20} in \eqref{eq2},\eqref{eq4}, and \eqref{eq5} produces the following homogeneous linear system of equations
\begin{equation}
\begin{cases}
\;(a_3 - \omega^2) \xi_0^1 + i\omega a_1 \xi_0^3 + i\omega a_2 \xi_0^4 = 0, \\[4pt]
\;i\omega b \xi_0^1 - \omega^2 \xi_0^3 = 0, \\[4pt]
\;i\omega c \xi_0^1 - \omega^2 \xi_0^4 = 0,
\end{cases}
\label{eq:system}
\end{equation}
Non-trivial solutions require the determinant of the coefficient matrix to vanish, leading to two distinct frequency solutions
\begin{eqnarray}
\omega^2 &=& 0,    \label{eq:trivial}\\
\omega^2 &=& \frac{m}{r^4}\frac{(1 + q) f^{q-1} g^{q(2+q)}}{\big(r-m(3 + 2q)\big)} \nonumber\\
&\times& \Big[ 2m^2(6 + 7q + 2q^2) 
 - 2m(4 + 3q)r + r^2 \Big].\label{eq:smallfreq}
\end{eqnarray}

The solution \eqref{eq:smallfreq} emerges from the compatibility condition of the system \eqref{eq:system}, describing stable oscillations constrained by the parameters $m$ and $q$. Also, Eqs.~\eqref{bigfreq} and ~\eqref{eq:smallfreq} represent the orbital and radial frequencies, respectively, as derived in \cite{toshmatov2019harmonic}. However, in Shirokov's approach, according to the assumptions stated herein, the frequency in Eq.~\eqref{eq:smallfreq} is not explicitly specified and is instead approximated uniformly for all three components: \( r \), \( \phi \), and \( t \).

\begin{figure}
    \centering
    \includegraphics[width=1\linewidth]{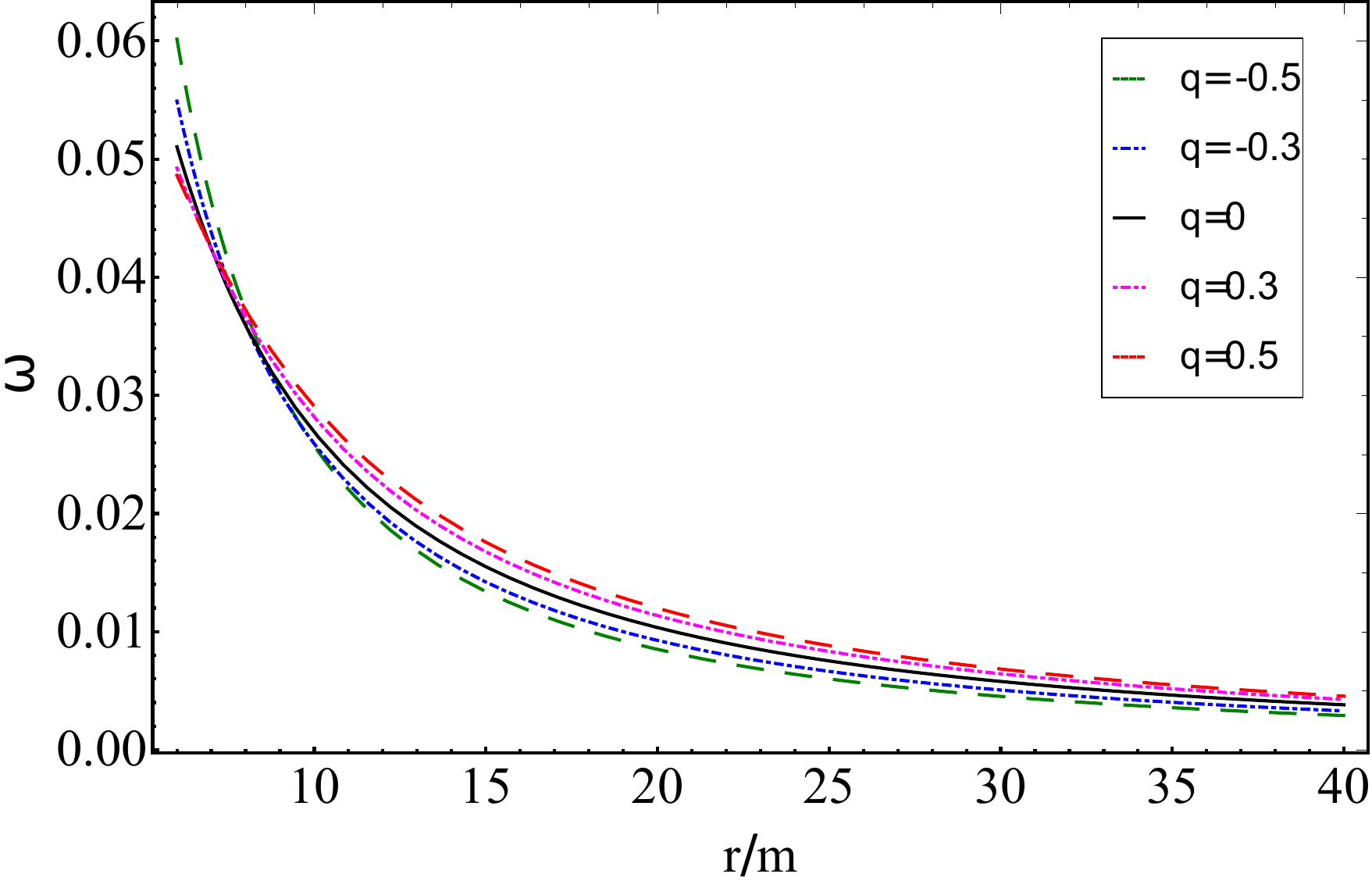}
    \caption{Dimensionless longitudinal oscillation frequency $\tilde{\omega}=m\omega$ as a function of $r/m$ for different values of $q$.}
    \label{fig:w1}
\end{figure}

\begin{figure}
    \centering
    \includegraphics[width=1\linewidth]{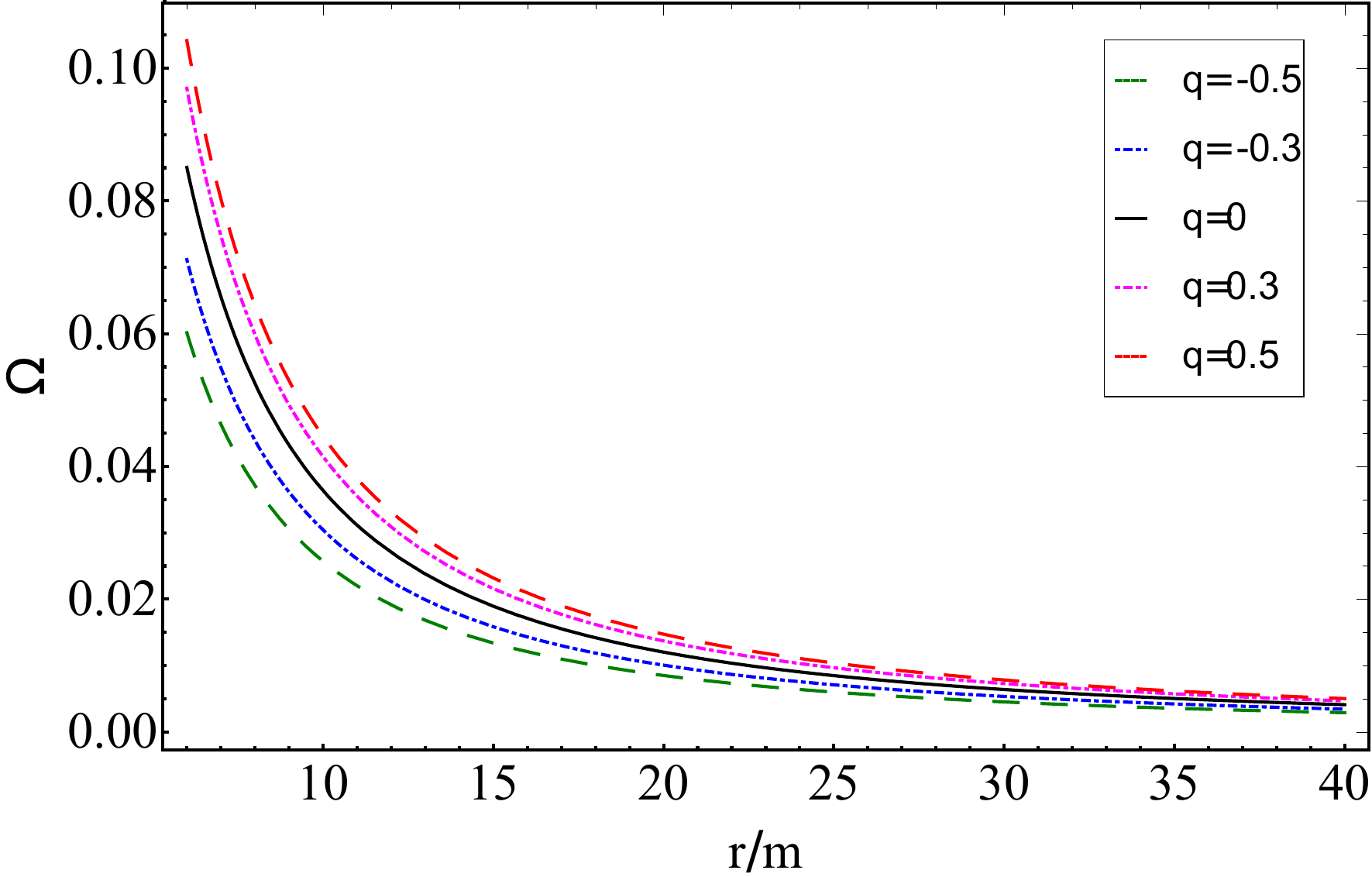}
    \caption{Dimensionless transverse oscillation frequency $\Tilde{\Omega}=m\Omega$ as a function of $r/m$ for different values of $q$.}
    \label{fig:w2}
\end{figure}

As shown in Fig. \ref{fig:w1} and Fig. \ref{fig:w2}, the quadrupole parameter affects the dynamics of both longitudinal and transverse oscillations, in particular small oscillations vanish with increasing distance $r$. At short distances, in the case of  prolate deformations ($q>0 $), the transverse oscillation frequency increases. Similarly, for oblate deformations ($q < 0$), the longitudinal oscillation frequency also reaches a significant level.

To obtain the oscillation periods, we first perform a Taylor expansion of \(\Omega\) and \(\omega\) with respect to the deformation parameter \(q\), under the assumption that \(q\) is small. These expanded forms are then substituted into the formulas for the oscillation periods $T_{\theta}=2\pi/\Omega$ and $T_{r}=T_{\phi}=2\pi/ \omega$. In order to explore the impact of the quadrupole moment $q$ on the orbital period of test particles in the Zipoy-Voorhees spacetime, we recover the total mass of the source as $M=m(1+q)$ and subsequently, expand the resulting expressions as series in inverse powers of the radial coordinate $(1/r)$, retaining linear terms in $q$, which allows a direct analysis of relativistic and quadrupolar corrections to the oscillation cyclic frequencies and periods at large distances from the central mass 
\begin{eqnarray}\label{freqns}
    \Omega&\approx&\sqrt{\frac{M}{r^3}}\left(1+\frac{3M(1-q)}{2r}+\frac{3M^2(9-10q)}{8r^2}\right) ,\\
    \omega&\approx&\sqrt{\frac{M}{r^3}}\left(1-\frac{3M(1+q)}{2r}-\frac{M^2(45-14q)}{8r^2}\right) ,
\end{eqnarray}
\begin{eqnarray}\label{periodr}
T_{\theta}&\approx&T_{0}\left(1-\frac{3M(1-q)}{2r}-\frac{3M^2(3+2q)}{8r^2}\right),\\
T_{r}&\approx&T_{0}\left(1+\frac{3M(1+q)}{2r}+\frac{M^2(63+22q)}{8r^2}\right),
\end{eqnarray}
where $T_{0}=2\pi/\sqrt{\frac{M}{r^{3}}}$ is the Newtonian period.  The difference between $T_{\theta}$ and $T_{r}$ (or $T_{\phi}$) is defined as
\begin{equation}\label{eq24}
\begin{gathered}
\Delta T=T_{\theta}-T_{r}\approx T_0\left(-\frac{3M}{r}-\frac{M^2(18+7q)}{2r^2}\right)
\end{gathered}
\end{equation}
Here we see that unlike the Schwarzschild spacetime, the contribution of $q$ appears in the subsequent term $\sim M^2/r^2$, which requires high precision techniques to measure the effects of $q$.
\begin{figure*}
{\hfill
\includegraphics[width=8cm]{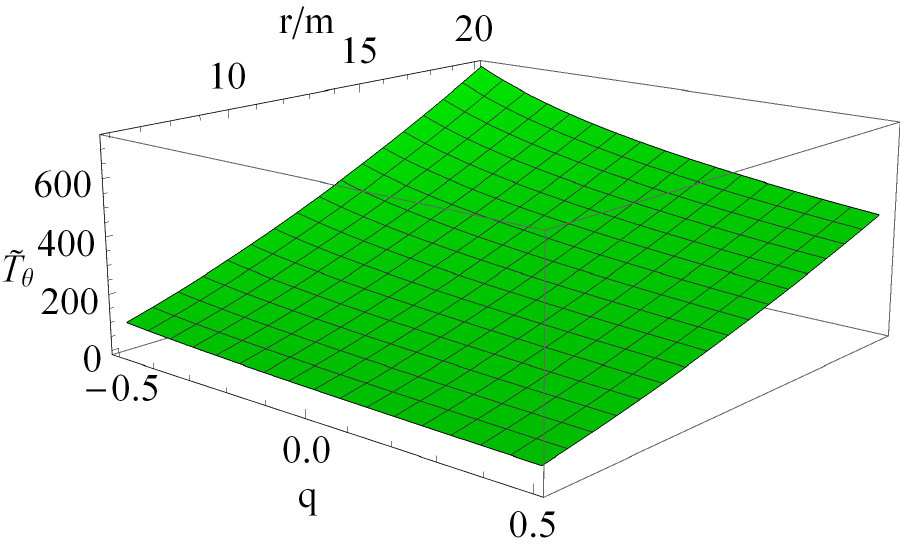}\hfill
\includegraphics[width=8cm]{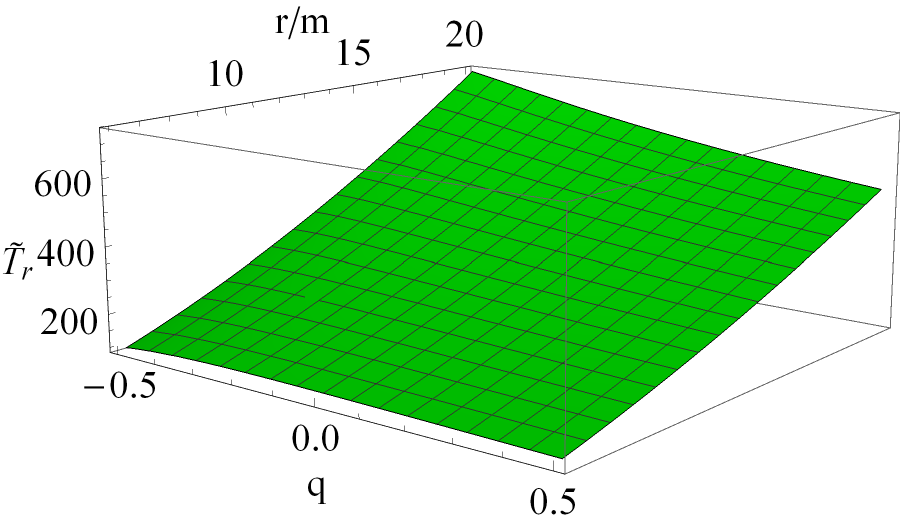}\hfill} 

{\hfill
\includegraphics[width=8cm]{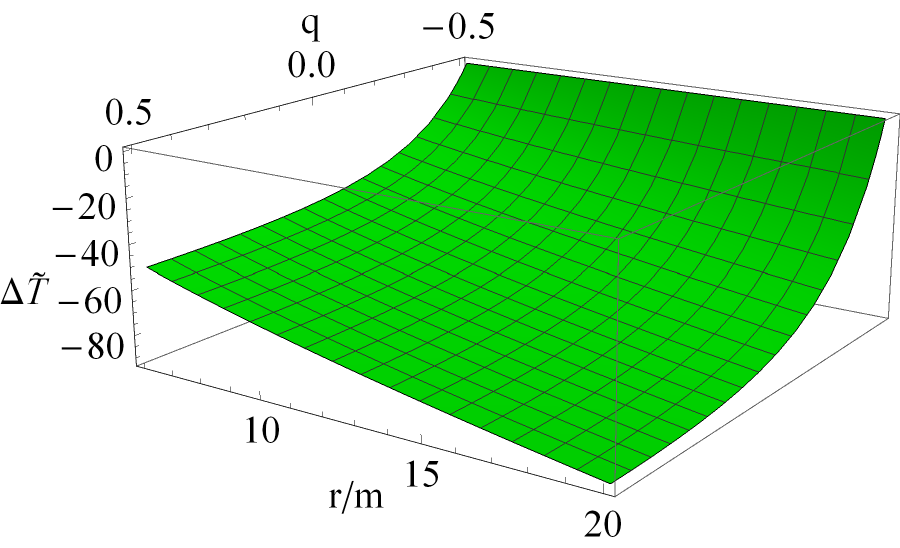}\hfill
} 

\caption{
Dependence the periods  $\tilde{T_{\theta}}$, $\tilde{T_{r}}$  and $\Delta \tilde{T}$ on  $q$ and $r/m$.}
\label{fig:3Dplot} 
\end{figure*}

Eqs. \eqref{eq:smallfreq} and \eqref{bigfreq} incorporate the influence of the $q$ parameter in the framework of Einstein's theory of gravitation. 
Fig. \ref{fig:3Dplot} shows the dependence of $\tilde{T_{\theta}}$, $\tilde{T_{r}}$, and $\Delta \tilde{T}$ on $q$ and $r/m$ . $\tilde{T_{\theta}}$, $\tilde{T_{r}}$, and $\Delta \tilde{T}$ represent $T_{\theta}/m$, $T_{r}/m$, and
 $\Delta T/m$, respectively. An increase in the value of $q$ produces  a decrease in the value of $T_{r}$ and a decrease in the value of $T_{\theta}$. In the limiting case ${q\rightarrow 0}$, we recover the values for the Schwarzschild solution. 
 
\begin{figure*}[htbp]
{\hfill
\includegraphics[width=17cm]{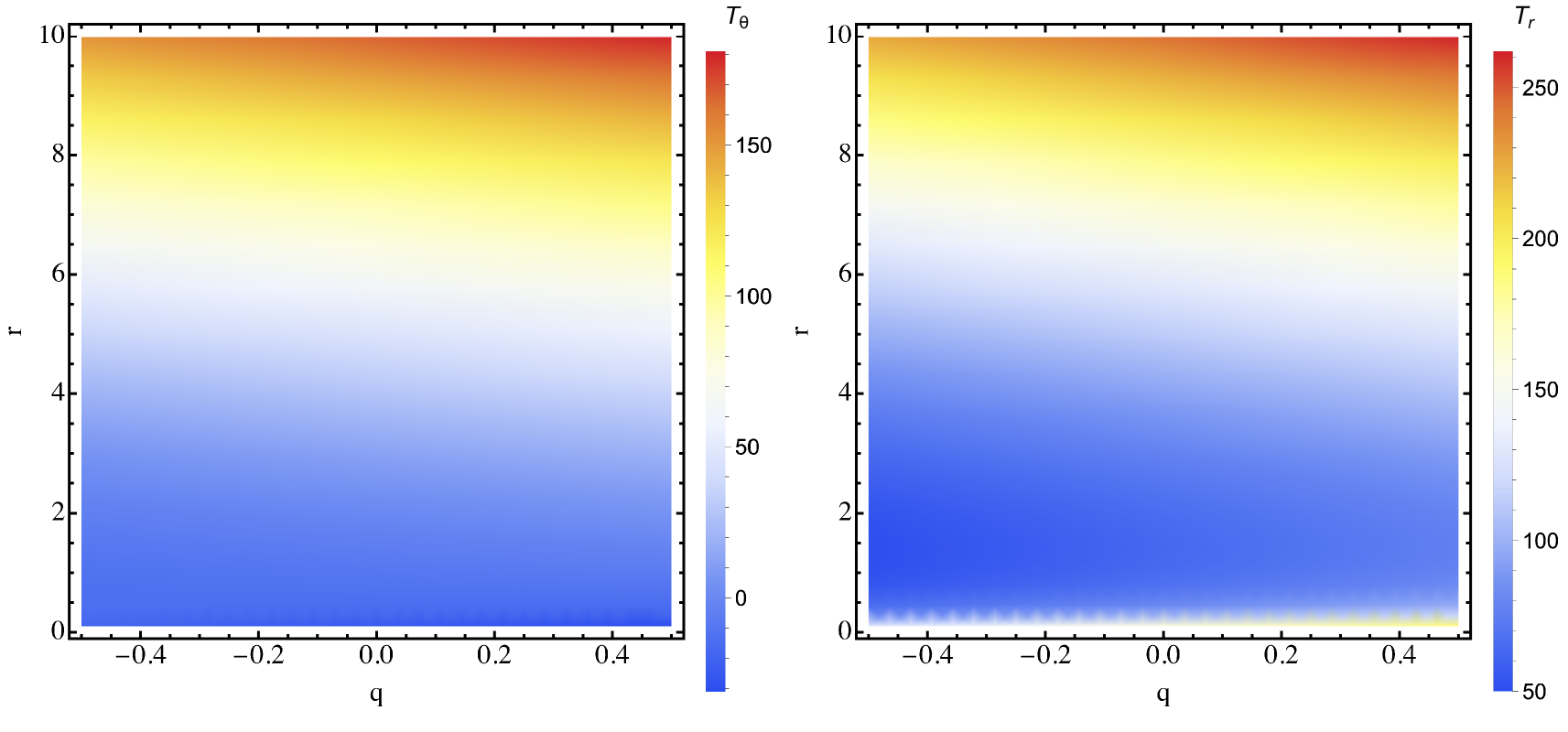}
\hfill} 
\caption{The same as in Fig. \ref{fig:3Dplot} but here as a heatmap showing the dependence of the orbital periods  $T_{\theta}$, $T_{r}$ from Eq.\eqref{periodr} on  $q$  and $r$, $M=1$.}
\label{fig:heatmap}
\end{figure*}

 Furthermore, an alternative representation of the periods \eqref{periodr} can be examined using the heat map in Fig. \ref{fig:heatmap}, where it is evident that both \(T_\theta\) and \(T_r = T_\phi\) increase monotonically with the radial coordinate \(r\). The left panel shows that the polar period \(T_\theta\) is also sensitive to the quadrupole parameter \(q\), with larger negative values of \(q\) leading to significantly higher values of \(T_\theta\) at large radii. In contrast, the right panel indicates that the radial and azimuthal periods are predominantly dependent on \(r\) and exhibit a weaker dependence on \(q\). Moreover, for prolate sources ($q<0$), the radial and azimuthal periods tend to be negative. This visual representation highlights how the deformation of the central object (encoded in \(q\)) affects the orbital motion more strongly than the radial and azimuthal components.

To obtain the deviation equations for $\xi^3$ and $\xi^4$, we substitute the values of $b$ and $c$ in Eqs. \eqref{eq4} and \eqref{eq5}, and obtain 
\begin{eqnarray}\label{eq25}
\xi_{0}^{3}&=&\frac{2}{r}\left(1+\frac{M(3-q)}{r}\right) e^{i \pi/2}\xi_{0}^{1},\\
\xi_{0}^{4}&=&2\sqrt{\frac{M}{r}}\left(1+\frac{M(10-q)}{2r}\right)e^{i \pi/2}\xi_{0}^{1}.
\end{eqnarray}
Taking into account the  results obtained above, Eqs. \eqref{eq19} and \eqref{eq20} can be written in the following form
\begin{align}\label{allvibrations} \nonumber
\xi^{1} & =\xi_{0}^{1} \sin{\omega s}, \nonumber\\
\xi^{2} & =\xi_{0}^{2}\sin{\Omega s}, \nonumber \\
\xi^{3} & =\frac{2}{r}\left(1+\frac{M(3-q)}{r}\right) e^{i \pi/2}\xi_{0}^{1}\cos{\omega s},\\ 
\xi^{4} & =2\sqrt{\frac{M}{r}}\left(1+\frac{M(10-q)}{2r}\right)e^{i \pi/2}\xi_{0}^{1}\cos{\omega s}. \nonumber
\end{align}

From Eqs. \eqref{allvibrations}, one can clearly see that the vibrations of the test particles inside a satellite consist of oscillations in and out of the plane. For in-plane oscillations described by $\xi^1$, we observe a deviation within the orbital plane (radial direction). In addition, for $\xi^3$ we note  another deviation within the orbital plane (orthogonal to 
$\xi^1$, e.g., tangential). In particular, the combination of $\xi^1$ and $\xi^3$ exhibits oscillations with a phase difference, resulting in an elliptical path in the $(\xi^1,\xi^3)$ plane. Out-of-plane (vertical $\theta$) oscillations are represented solely by  $\xi^2$,  which is perpendicular to the orbital plane and takes the form of harmonic oscillations. The frequency $\Omega$ differs from $\omega$,  resulting in a different period. The time component $\xi^4$ is not a spatial displacement, but indicates how the proper time of the center of inertia of the satellite and that of the particle varies and is related to the relativistic time dilation effects of the geodesic deviation.

For clarity, we consider a model in which a satellite, like the Moon, both orbits Earth and rotates about $\xi^2$, with a rotational period equal to its orbital period, according to Eq. \eqref{u4}. 
\begin{eqnarray}
{\cal T} &=& T_{0}\sqrt{f^{-(q+1)}\frac{r-m(2+q)}{r-m(3+2q)}}\nonumber\\
&\approx& T_0\left(1+\frac{3M}{2r}+\frac{3M^2(9-4q)}{8r^2}\right) .    
\end{eqnarray}\label{eq27}
Inside the satellite there are test particles that experience tiny free oscillations, and the satellite carries instruments that can track  with high precision the motion of particles $\xi^1$ and $\xi^2$ in the  $(\xi^1,\xi^2)$ - plane.

We wish to find out how far a test particle shifts in the 
$\xi^2$ direction (orthogonal to the orbital plane) after completing 
$n$ oscillations along the 
$\xi^1$-axis (in-plane direction). The idea is that due to relativistic effects, the periods of oscillation in $\xi^1$ and $\xi^2$ 
differ as discussed above, so even though they start at the same point, they get out of phase over time. This leads to a measurable cumulative shift in the $\xi^2$-direction after $n$ cycles of oscillation in 
$\xi^1$.

Based on (\ref{eq24} and \ref{allvibrations}), if the number of completed oscillations $n= (1,2,3..)$  is relatively small, then the vertical displacement becomes
\begin{eqnarray} \label{vertshift}
 \xi^{2}&=&\xi_{0}^{2}\sin{\frac{2 \pi n T_{r}}{T_{\theta}}}\nonumber\\&\approx&\xi_{0}^{2}\sin\left[2 \pi n\left(1+\frac{3M}{r}+\frac{M^2(27-2q)}{2r^2}\right)\right]\\
&\approx& \xi_{0}^{2} \frac{6\pi n M}{r}\left(1+\frac{M(27-2q)}{6r}\right)\nonumber .   
\end{eqnarray}
Here we see again, that the contribution due to $q$ appears in the second post-Newtonian term $\sim (M/r)^2$.

\begin{figure}
    \centering    \includegraphics[width=1\linewidth]{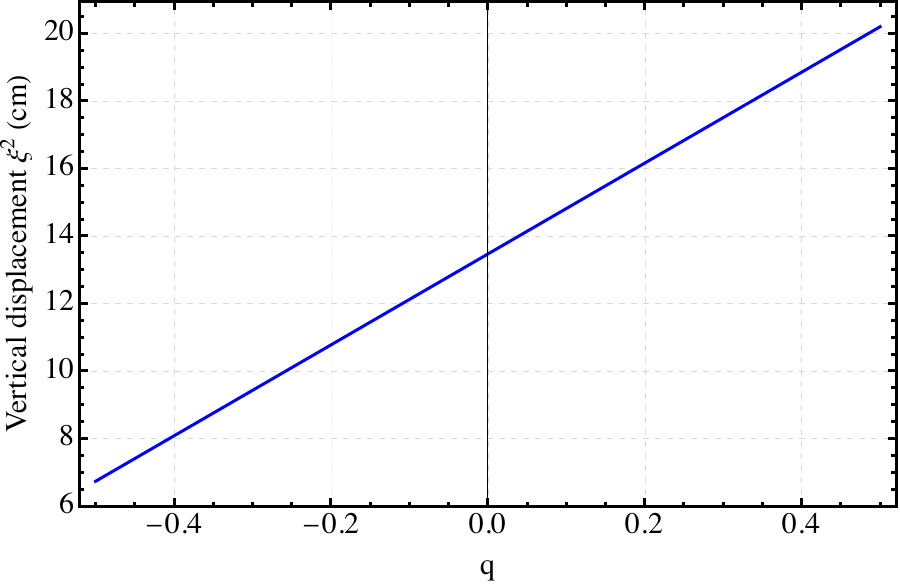}
    \caption{Dependence of the vertical shift $\xi^2$, Eq. \eqref{vertshift}, scaled in powers of $10^{-6}$   on the deformation parameter $q$. 
        The following values are used: $\xi_0^2 = 10~\text{cm}$,  
        $n = 10$, $m = 0.5~\text{cm}$ (gravitational radius), and 
        $r = 7 \times 10^8~\text{cm}$.}
    \label{fig:vertdisp}
\end{figure}

One can recover the values for the Schwarzschild metric for $q \to 0$,  where the vertical shift $\xi^2$ is of the order of $10^{-6}$ cm. In the case of the $q-$metric, from Fig.\ref{fig:vertdisp}, the orbital shift increases for $q > -0.9$. Moreover, as the source becomes more oblate, 
the orbital shift decreases. Such magnitudes can be detected by satellite-based interferometry or techniques like laser interferometry (as in LISA \cite{amaro2017laser} or GRACE Follow-On \cite{abich2019orbit}), which allow for distance measurements between test masses with accuracy down to scales of $10^{-9}$m or even $10^{-12}$m, making them highly suitable for detecting deviations such as those predicted by the Shirokov effect. 

\section{Shapiro time delay}
\label{sec:shap}

Another way to test gravity is by measuring the time delay of light waves passing near a massive object, an effect caused by the spacetime curvature and also known as the fourth test of GR, studied by I. Shapiro in \cite{shapiro1964fourth}. Specifically, Shapiro examined the Schwarzschild spacetime, which describes the gravitational field of the Sun, neglecting Earth's motion between the pulse transmission and echo reception. To calculate the time delay in the spacetime described by  the $q$-metric \eqref{metric}, we follow  the approach outlined in \cite{bambi2018introduction}.
For massless particles, one can write for \eqref{metric}
\begin{equation} \label{consmass}
f^{-1 - q} g^{q (2 + q)} \, \dot r^2 - f^{1 + q} \, \dot t^2 + f^{-q} r^2 \, \dot \phi^2 = 0.
\end{equation}
Then, the conserved energy $E$ and the angular momentum $L$ \cite{2025GReGr..57...91M} are
\begin{equation} \label{consEL}
\dot{t} = -\frac{E}{f^{1 + q}}, \quad
\dot{\phi} = \frac{L}{f^{-q} r^2}.
\end{equation}
Furthermore,  to remove affine parameter $s$, we use the relationship 
\begin{equation}
\frac{dr}{d s} = \frac{dr}{dt} \frac{dt}{ds} = - \frac{dr}{dt} \frac{E}{f^{1 + q}}.
\end{equation}
At $r = r_C$ , we have $dr/dt= 0$, and combining \eqref{consEL} with \eqref{consmass}, we find
\begin{equation}
L^2 = E^2 r_C^2 f(r_C)^{-1 - 2q}  .
\label{LE}
\end{equation}
Replacing the expressions  \eqref{consEL} and \eqref{LE} in the general equation \eqref{consmass}, we finally obtain 

\begin{equation}
\left(\frac{dr}{dt}\right)^2 \frac{g^{q(2 + q)}}{f^{3(1 + q)}}  - \frac{1}{f^{1+q}} + \frac{f(r_C)^{-1 - 2q}  r_C^2}{f^{-q} r^2} = 0
\end{equation}
or
\begin{equation}
\frac{dr}{dt}=\left[ f^{2 + 2q} g^{-q(2 + q)} \left(1 - \frac{f^{1 + 2q}  r_C^2}{f_C^{1 + 2q} r^2} \right) \right]^{1/2}
\end{equation}

\begin{figure}
    \centering
    \includegraphics[width=0.6\linewidth]{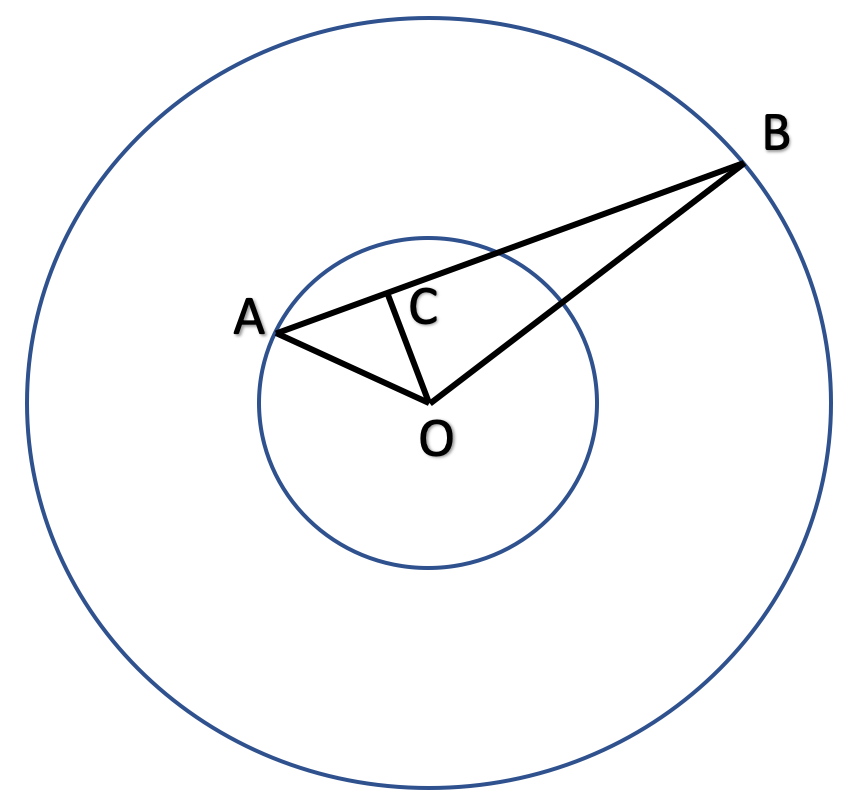}
    \caption{Schematic illustration of Shapiro time delay. The Sun is at the central point O. Two planets located at A and B move along circular orbits around the Sun. C is the closest point to the Sun for a ray propagating from A to B.}
    \label{fig:scheme}
\end{figure}

\begin{figure}
    \centering
    \includegraphics[width=0.99\linewidth]{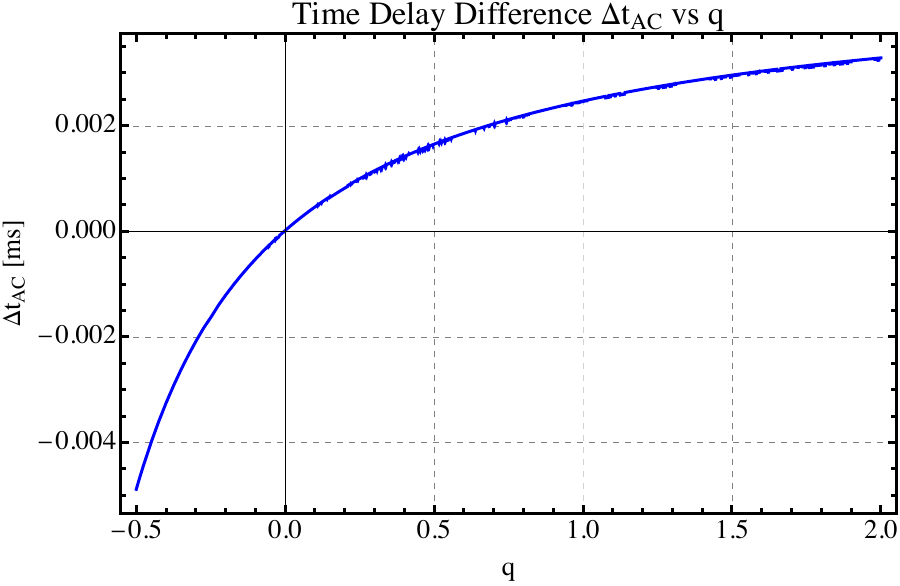}
    \caption{Numerical integration of the difference of time delay $\Delta t_{AC} = t_{AC} [q]- t_{AC} [0]$, according to Eq. \eqref{integral},  as a function of the deformation parameter $q$.}
    \label{fig:diffinq}
\end{figure}

The time it takes for a light ray signal to go from A to C in Fig.~\ref{fig:scheme} is 

\begin{equation} \label{integral}
t_{AC} = \frac{1}{c} \int_{r_C}^{r_A} \frac{dr}{\sqrt{\left[ 1 - \frac{f(r)^{1+2q}}{f(r_C)^{1+2q}} \frac{r_C^2}{r^2} \right] \frac{f(r)^{2(1+q)}}{ g(r)^{q(2+q)}}}}.
\end{equation}
The numerical integration of \eqref{integral} is shown in Fig.~\ref{fig:diffinq}, where it is evident that the time delay depends on the quadrupole parameter $q$. If we consider only the first-order terms in the expansion in $m$ and $q$, we obtain
\begin{align}
& \left[ 1 - \frac{f(r)^{1+2q}}{f(r_C)^{1+2q}} \frac{r_C^2}{r^2} \right] 
\frac{f(r)^{2(1+q)}}{ g(r)^{q(2+q)}}  \\ \nonumber
& = \left(1 - \frac{2m}{r}\right)^{2(1 + q)} 
\left(1 - \frac{m^2}{2mr - r^2}\right)^{-q(2 + q)} \\ \nonumber
& \quad \times  \left[1 - \frac{ \left(1 - \frac{2m}{r} \right)^{1 + 2q} 
\left(1 - \frac{2m}{r_C} \right)^{-1 - 2q} r_C^2 }{r^2} \right] \\ \nonumber
& \approx 1 - \frac{r_C^2}{r^2} 
- \frac{2m (r - r_C) \left(2(1 + q)r + (3 + 4q)r_C\right)}{r^3}
\end{align}
and therefore
\begin{align}
&\frac{1}{\sqrt{\left[ 1 - \frac{f(r)^{1+2q}}{f(r_C)^{1+2q}} \frac{r_C^2}{r^2} \right] 
\frac{f(r)^{2(1+q)}}{ g(r)^{q(2+q)}}}} 
\\ \nonumber
& \approx  \frac{r}{\sqrt{r^2 - r_{C}^2}}
+ \frac{r_s \left[2 (1 + q) r^2 + (1 + 2q) r r_C - (3 + 4 q) r_C^2\right]}{2 \left(r^2 - r_C^2 \right)^{3/2}}.
\end{align}

Eventually, the time that it takes for the electromagnetic signal to go from point A to point C turns out to be
\begin{align} \label{inttime}
t_{AC} &= \frac{1}{c} \sqrt{r_A^2 - r_C^2} 
+ \frac{(1+q)r_s}{c} \ln \left( \frac{r_A + \sqrt{r_A^2 - r_C^2}}{r_C} \right)+ \notag \\
&\quad + \frac{(1+2q) r_s}{2c} \sqrt{ \frac{r_A - r_C}{r_A + r_C}} 
\end{align}
where $c$ is the speed of light.
In the case of a flat spacetime, the time that the electromagnetic signal would take to go from point \(A\) to point \(C\) is
\begin{equation}
\tilde{t}_{AC}=\frac{1}{c}\sqrt{r_{A}^{2}-r_{C}^{2}}\,,
\end{equation}
and corresponds to the leading order term in \eqref{inttime}. The total time that the electromagnetic signal takes to go from \(A\) to \(B\) and come back to \(A\) is
\begin{equation}
t_{\mathrm{tot}}=2t_{AC}+2t_{BC}\,,
\end{equation}
while in flat spacetime it would be
\begin{equation}
\tilde{t}_{\mathrm{tot}}=2\tilde{t}_{AC}+2\tilde{t}_{BC}=\frac{2}{c}\sqrt{r_{A}^{2}-r_{C}^{2}}+\frac{2}{c}\sqrt{r_{B}^{2}-r_{C}^{2}}\,.
\end{equation}
The maximum time delay with respect to the flat spacetime $\delta t_{\mathrm{max}}  =\tilde{t}_{\mathrm{tot}}-t_{\mathrm{tot}}$ is when \(r_{C}=R_{\odot}\), where \(R_{\odot}\) is the radius of the surface of the Sun, $r_s = 2G m/c^2$. The result is (\(R_{\odot}\ll r_{A},r_{B}\))
\begin{align} \label{tmaxvsq}
\delta t_{\mathrm{max}} & =\frac{2 r_s}{c} \left[1 +2q+ (1+q) \ln\left(\frac{4r_{A}r_{B}}{R_{\odot}^{2}}\right) \right]\,\nonumber\\
&=\frac{2 R_s}{c} \left[1+q+ \ln\left(\frac{4r_{A}r_{B}}{R_{\odot}^{2}}\right) \right]
\end{align}
where $R_s=2GM/c^2$ and $M=m(1+q)$ is the total mass of the source of gravity as earlier.
One can recover the time delay for the Schwarzschild spacetime for $q=0$ in Eq.\eqref{tmaxvsq}.
\begin{figure}
    \centering    \includegraphics[width=0.9\linewidth]{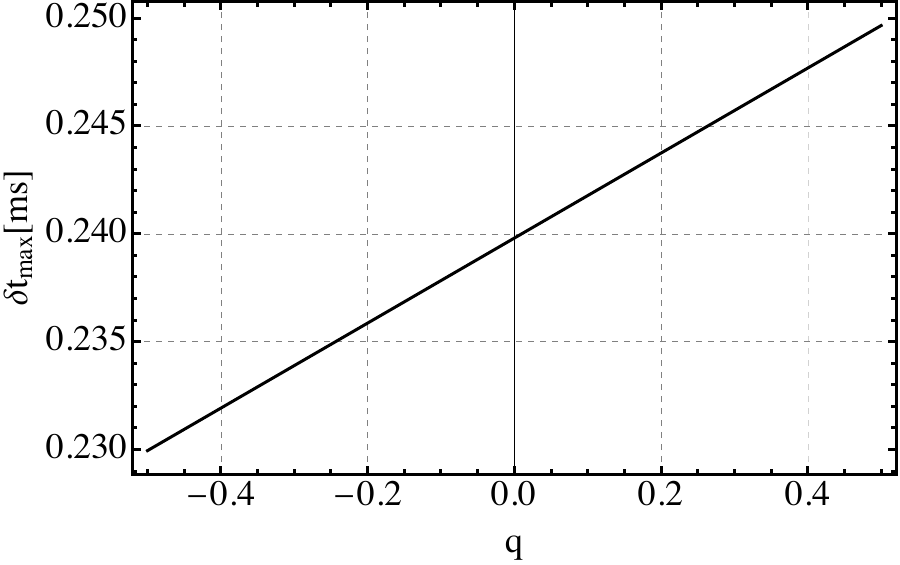}
    \caption{The time delay Eq.\eqref{tmaxvsq} with respect to the quadrupole parameter when a radar signal travels
back and forth through the path close to the Sun, when the exterior field is described by the $q$-metric.}
    \label{fig:shapirotd}
\end{figure}

From Fig. \ref{fig:shapirotd}, it is evident that for more oblate compact object ($q>0$), the time delay is greater compared to the Schwarzschild case ($q=0$), where the time delay corresponds to $\delta t \approx 0.24 ms$. The propagation of a light signal is subject to an increased time delay when the compact object undergoes a oblate deformation along the light's trajectory.
\\
\section{Conclusions}
\label{sec:con}

In this work, we have explored the properties of the $q$-metric, focusing on two key relativistic effects: the Shirokov effect and the Shapiro time delay. Our analysis demonstrates that the geodesic deviation in the $q$-metric leads to oscillatory behavior influenced by the quadrupole parameter, distinguishing it from the Schwarzschild case. In particular, the period $T_{r}$ equals the orbital period of a test body in a circular orbit of radius $r$ in the proper (rotating) reference frame. These oscillations of the test body relative to the satellite's center are caused by the small angle of inclination between the circular orbits of the test body and the satellite's center of inertia. However, oscillations in the in-plane orbit are due to the quasi-elliptical nature of the test body's orbit with a general relativistic perihelion shift similar to that observed in Mercury.

Additionally, we have derived the Shapiro time delay in the $q$-metric and found that its magnitude depends on the quadrupole deformation of the spacetime. We conclude that the main result presented in \cite{Chakrabarty:2022fbd} is counterintuitive.

These results provide further insight into the role of spacetime deformations in gravitational lensing and orbital dynamics. The findings could be relevant for astrophysical applications, particularly in the study of compact objects with significant quadrupole moments. Future observational tests, including the precision timing of signals from pulsars and spacecraft, may offer further constraints on deviations from spherical symmetry in strong gravitational fields.
\\

\section*{Acknowledgements}

 DU is supported by Grant No. AP22682939; KB and AU are supported by Grant No. AP19680128, AI is supported by Grant No. AP14870501, all from the Science Committee of the Ministry of Science and Higher Education of the Republic of Kazakhstan. The work of HQ was supported by PAPIIT-DGAPA-UNAM, grant No. 108225, and by CONAHCYT,  grant No. CBF-2025-I-253. AI is
grateful for the kind hospitality of the Rudolf Peierls Centre
for Theoretical Physics at the University of Oxford, where some work on this research was done, and extends appreciation to the
Yessenov Foundation for its financial support.



\bibliography{0refs}
\end{document}